\documentstyle[twocolumn,prl,aps,epsf,floats]{revtex}

\begin{document}
\draft


\wideabs{

\title{Reorientation of Spin Density Waves in Cr(001) Films 
induced by Fe(001) Cap Layers}

\author{P.~B\"odeker$^{(1)}$, A.~Hucht$^{(2)}$, A.~Schreyer$^{(1,3)}$,
J.~Borchers$^{(3)}$, F.~G\"uthoff$^{(4)}$, and H.~Zabel$^{(1)}$}

\address{$^{(1)}$Institut f\"ur Experimentalphysik / Festk\"orperphysik,
Ruhr-Universit\"at Bochum, D-44780 Bochum, Germany \\
$^{(2)}$Theoretische Physik, Gerhard-Mercator-Universit\"at Duisburg,
D-47048 Duisburg, Germany \\
$^{(3)}$National Institute of Standards and Technology, Gaithersburg,
Maryland 20899, USA \\
$^{(4)}$Forschungszentrum J\"ulich, D-52425 J\"ulich, Germany }


\date{Received November 17, 1997}
\maketitle

\begin{abstract}
Proximity effects of 20\,\AA\
thin Fe layers on the spin density waves (SDWs) in epitaxial Cr(001) films
are revealed by neutron scattering.
Unlike in bulk Cr we observe a SDW with its wave vector ${\bf Q}$
pointing along only one \{100\} direction which depends dramatically on the
film thickness $t_{\mathrm{Cr}}$.
For $t_{\mathrm{Cr}}< 250$\,\AA\
the SDW propagates out-of-plane with the spins in the film plane.
For $t_{\mathrm{Cr}}>1000$\,\AA\ the SDW propagates in the film plane with the 
spins out-of-plane
perpendicular to the in-plane Fe moments.
This reorientation
transition is explained by frustration effects in the
antiferromagnetic interaction between Fe and Cr across the
Fe/Cr interface due to steps at the interface.
\end{abstract}


}
\setcounter{page}{914}
\markright{\rm{\it Physical Review Letters} {\bf 81}, No.~4 (1998)}
\thispagestyle{myheadings}
\pagestyle{myheadings}


While the incommensurate spin density wave (SDW) antiferromagnetism is well
established for bulk Cr \cite {Fawcett},
it is presently of high interest to analyze how the magnetic properties of
Cr are altered either by reduced dimensionality in thin films or by
proximity effects to ferromagnetic (FM) layers.
The magnetic state of Cr is particularly interesting since
ultrathin Cr films play an important role in exchange coupled Fe/Cr
superlattices exhibiting giant magneto-resistance effects
\cite{Grunberg,Fert}. Also for theoretical treatments of the exchange
coupling it is uncertain whether the Cr
spacer layer should be treated as a paramagnet, an antiferromagnet, or as a
proximity induced antiferromagnet \cite{Stiles}. In this context, the
role of the Fe/Cr interface is a matter of intense study
\cite{FeCrexp,Stoeffler}. 
Magnetic domain imaging of an
Fe layer deposited on a wedge shaped Cr layer on an Fe whisker shows a domain
pattern switching between parallel and anti-parallel
alignments having a periodicity of two Cr(001) monolayers
and a phase shift consistent with a SDW state \cite{Ung}.
More recently, neutron scattering and perturbed angular correlation
spectrocopy (PACS) have been used on Fe/Cr(001) superlattices to
investigate the magnetic structure of Cr directly
for Cr film thicknesses $t_{\mathrm{Cr}}$ of about 30 - 400 \AA \
\cite {Meersschaut,Fullerton,Schreyer}.
Although some inconsistencies still remain, these experiments show that the
SDW state collapses for Cr films well below the period
$\Lambda$ of the SDW.

The aim of the present
work is to gain a basic understanding of the effect of FM proximity
layers on the  magnetic properties of
thin Cr(001) films in the SDW phase. The thickness range of
the Cr films (200 - 3000 \AA) is chosen such that the question of the
presence of a SDW state is not an issue. Using neutron scattering we find
that the propagation direction of
the SDW depends dramatically on the Cr film thickness.
Our experimental results are rationalized by computer
simulations using a Heisenberg model which takes realistic Fe/Cr interfaces
with interfacial roughness and interdiffusion into account. Complementary
experiments with synchrotron radiation will be discussed elsewhere.

We have grown epitaxial Fe/Cr(001) bilayers by molecular beam epitaxy
on Al$_2$O$_3$(1$\bar1$02) substrates with a 500\,\AA\ thick Nb(001)
buffer layer, following well established
growth recipies \cite{Dur,KB}.
Cr(001) films with thicknesses from 200\,-\,3000\, \AA \ were grown on the Nb
buffer layer at 450$^\circ$C with a growth rate of 0.1\,\AA/s.
As evidenced by reflection high energy electron diffraction (RHEED) during
growth, the crystalline quality of the Cr near the Cr/Nb interface is not
very high due to the 14 \% lattice mismatch between Nb and Cr.
However, with growing Cr thickness the film quality improves dramatically
\cite{KB}.
After annealing for 30\,min at 750$^\circ$ the Fe cap layer
was grown at 300$^\circ$ at a rate of 0.1\,\AA/s.
For the scattering experiments it was necessary to keep the
absolute amount of Cr in the samples roughly constant. Therefore,
for samples with $t_{\mathrm{Cr}}<1000$\,\AA\ the Fe/Cr
structure was repeated several times up to a minimum total $t_{\mathrm{Cr}}$ of
2000\,\AA\ .
All additional Fe/Cr layers were grown at 300$^\circ$C.
For protection against oxidation all samples were covered with
a 20\,\AA\ Cr layer \cite{Stierle}.
X-ray scattering shows that all samples are of high crystalline quality,
with an out-of-plane crystalline coherence length
of 60\,-\,80\% of the total film thickness and a mosaic spread of about
$0.2{^\circ}$ FWHM measured at the (002) peak.

To determine the magnetic structure neutron scattering
experiments were performed on the triple-axis spectrometers BT-2 of the
National Institute of Standards and Technology
and UNIDAS at the KFA J\"ulich, Germany. In both cases we used pyrolytic
Graphite PG(002) monochromator and analyzer crystals to select a wavelength
of $\lambda=2.351$\,\AA. Graphite filters suppressed any $\lambda/2$
contamination.

Bulk Cr exhibits an incommensurate SDW, i.e.~the magnitude of
the antiferromagnetically aligned Cr magnetic moments $\mu$
varies sinusoidally  with a temperature dependent period $\Lambda$ of
about 21 lattice constants at T = 0 \cite {Fawcett}.
The incommensurability is ascribed to a nesting vector along the \{100\}
directions of the Cr Fermi surface.
The wavevector $\vec {\bf Q}$
defines the direction of propagation of the SDW.
At lowest temperatures a longitudinal SDW (LSDW) forms,
i.e.~$\vec \mu$ is parallel to ${\vec {\bf  Q}}$.
Above the spin-flip transition temperature, $T_{SF} = 123$ K, $\vec \mu$ is
perpendicular to $\vec {\bf Q}$, forming a transverse SDW (TSDW).
Above $T_N = 311$ K bulk Cr is paramagnetic.

The incommensurate modulation of the
antiferromagnetic (AF) spin structure by the SDW causes {\it two}
satellite peaks to occur
around the \{1,0,0\} positions \cite{Fawcett}, e.g.
at (0,0,1$\pm \delta$), (1$\pm \delta$,0,0), and
(0,1$\pm \delta$,0) which can be investigated by neutron scattering.
Here $\delta=1-|\vec {\bf Q}|$ with $|\vec {\bf Q}|$ in reciprocal lattice
units, $\delta = a/\Lambda$, and the Cr lattice constant a.
In the first case, the position of the satellites indicates $\vec {\bf Q}$
being oriented out of the film plane, the two latter cases occur for either
direction of in-plane propagation.
In addition, the polarization of the SDW (i.e.~TSDW or LSDW)
can be obtained by making use of the
selection rule for magnetic neutron scattering.
It requires a component of the magnetization vector $\vec {\mu}$
to be perpendicular to the scattering vector $|\vec {\bf q}|=(4\pi/\lambda)
\sin \theta$ where $\theta$ is the scattering angle.
Thus, a longitudinal SDW propagating along L, i.e.~out-of-plane, will generate
no intensity at (0,0,1$\pm \delta$). However, satellites will occur at
(1,0,0$\pm \delta$). A commensurate AF (AF$_0$)
phase, on the other hand, will yield a {\it single} peak of purely magnetic
origin at the Cr\{001\} positions.
Thus, with neutron scattering we can uniquely determine AF$_0$ and
SDW magnetic order as well as SDW propagation and polarization.
For a more detailed discussion see e.g.~Ref.~\cite {Sonntag}.

\begin{figure}[t]
\epsfxsize=8.5cm
\epsffile{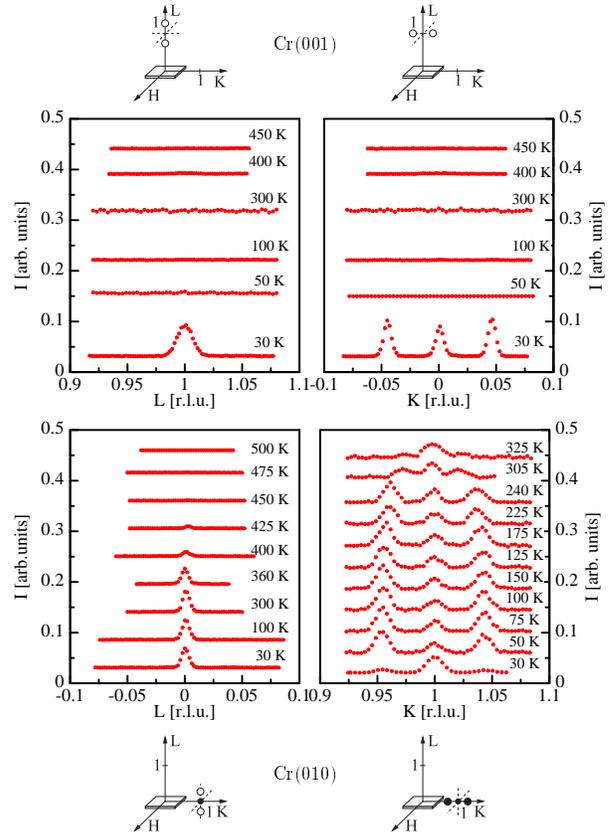}
\caption{
Neutron scans through the possible satellite positions
around the (001) (top) and (010) (bottom) position for a 3000\,\AA\ thick
Cr(001) film capped with a 20\,\AA\
Fe layer for 30\,K $\leq$ T $\leq$ 500\,K.
For each scan direction a schematic picture of the reciprocal lattice
with the open and solid circles as a representation of the
possible and observed satellites at 100 K is shown.}
\label{data}
\end{figure}

Fig.~\ref{data} reproduces neutron scans taken at possible satellite
positions around the Cr(001) and Cr(010) positions of a 3000\,\AA\
Cr(001) film capped with a 20\,\AA\
Fe layer for temperatures of 30\,-\,500\,K.
To make the spectra comparable, all intensities have been
normalized with respect to the structural reflections.
For all temperatures only satellites in scans along the in-plane K direction
occur, whereas in scans along L no satellites are found.
Thus, only SDWs propagating in the film plane are present.
For $T=30$\,K satellites appear around the (001) position.
From the selection rule described above we conclude that at
$T=30$\,K a LSDW propagates in the plane with the spins
dominantly oriented in-plane.
For temperatures $T\geq 50$\,K, however, we find the reversed situation.
The satellites occur only around the (010) position.
Thus, we conclude that the
SDW still propagates in-plane but that the polarization has changed
to transverse.
Moreover, the absence of any satellites in the K-scan around (001)
tells us that for temperatures $T\geq 50$\,K
we observe a TSDW with the spins
pointing out of the film plane, i.e.~perpendicular to the Fe/Cr interface.
The observed spin flip transition from the LSDW to the TSDW at approximately
$T_{\mathrm{SF}}^{\mathrm{film}}=(40\pm10)$\,K
occurs at a much lower temperature than in bulk
($T_{\mathrm{SF}}^{\mathrm{bulk}}=123$\,K).
Around $311$\,K the SDW satellites disappear consistent with
the bulk N\'eel temperature.

Another feature of the data of Fig.~\ref{data} is the presence of
intensity commensurate with the Cr(001) and Cr(010) positions
indicating an additional AF$_0$ phase.
From the temperature dependence the N\'eel temperature
is found to be $T^{\mathrm{com}} \approx 450$\,K.
Using the selection rule we again observe a flipping of the spins
from in-plane
below $T_{\mathrm{SF}}^{\mathrm{film}}=(40\pm10)$\,K to out-of-plane
above $T_{\mathrm{SF}}^{\mathrm{film}}$.
The origin of the AF$_0$ phase will be discussed below.

In summary we find
an in-plane LSDW and an AF$_0$ phase coexisting below 40\,K, both with spins
in-plane.
Above 40\,K an in-plane TSDW and an AF$_0$ phase coexist, both with spins
out-of-plane.
Thus, our measurements imply that the Cr moments are
oriented {\it perpendicular} to the Fe moments since magnetization
measurements confirmed that the Fe is magnetized in-plane.

\begin{figure}[t]
\epsfxsize=8cm
\epsffile{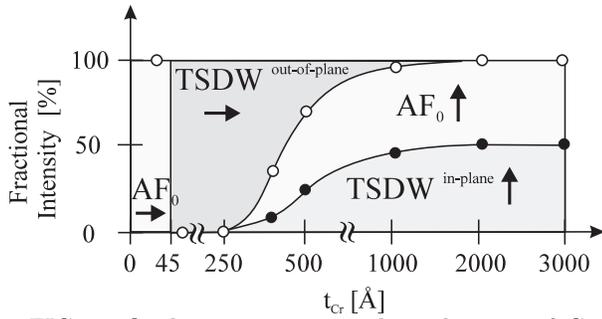}
\caption{Qualitative magnetic phase diagram of Cr films capped with 20\,\AA\
thin Fe layers as a function of the Cr thickness.
The superscripts describe the orientation of the
Q vector, the arrows the orientation of the Cr spins.
The solid and open
circles represent the relative intensities of the TSDW and the AF$_0$ peaks,
respectively.}
\label{phase2}
\end{figure}

In Fig.~\ref{phase2} results of equivalent measurements on a series of
samples with 250\,\AA\ $\leq$ $t_{\mathrm{Cr}}$ $\leq$ 3000\,\AA\ are summarized in
a qualitative phase diagram for $T=100$\,K.
Two points at 42 and 80\,\AA\ from earlier experiments by Schreyer et al.
\cite{Schreyer} have been added.
The diagram can be divided into four parts. For $t_{\mathrm{Cr}}$ $\leq$ 45\,\AA\ 
only an AF$_0$ phase with in-plane Cr spins exists. With increasing
$t_{\mathrm{Cr}}$ up to 250 \,\AA\ we observe an out-of-plane TSDW
with the Cr spins in-plane,
consistent with results of Fullerton et al.~\cite {Fullerton}.
For 250\,\AA\ $\leq$ $t_{\mathrm{Cr}}$ $\leq$ 1000\,\AA\
in- and out-of-plane TSDWs coexist.
Finally, for the thickest films, the reorientation to a TSDW propagating
in the film plane with spins out-of-plane is complete.
Interestingly, this reorientation is correlated with the
occurrence of a coexisting AF$_0$ phase with the same out-of-plane
spin orientation.

The observed reorientation effect
can be explained by considering a realistic Fe/Cr interface structure.
In the Fe/Cr system three different interactions are present:
a FM Fe-Fe and an AF Cr-Cr intra-layer
interaction within each Fe or Cr layer, and in addition an
AF \cite{Jun} inter-layer interaction between the Fe and
the Cr.
At an ideally flat interface all three interactions can co-exist without any 
frustration as long as all
moments are oriented in the film plane. However, at real interfaces
steps and interdiffusion may occur. Any step height of an uneven
number of Cr layers along the Fe/Cr interface
introduces frustrations between the Fe and Cr intra-layer interaction
on one hand, and the inter-layer interaction on the other hand 
\cite{BergerJMMM}.
It is not possible to minimize all three coupling energies
independently. Thus, the resulting spin structure
depends on the values of the respective coupling constants.
The following four limiting cases can be distinguished assuming a single
monoatomic step at the interface.
If the interface coupling is large compared to the Fe or Cr coupling constant,
a domain wall forms in the Fe (case 1) or in the Cr (case 2).
For a very small interface-coupling the ideal FM and AF order in
the Fe and Cr layers can be preserved by a domain wall forming along the
interface (case 3). If, however, the AF interface coupling is of intermediate
magnitude, the system can react by reorienting the Cr moments perpendicular
to the Fe (case 4).

\begin{figure}[t]
\epsfxsize=8.5cm
\epsffile{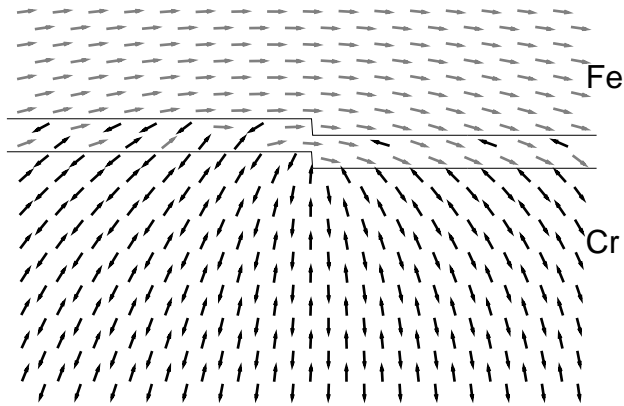}
\caption{
Ground state spin structure near an Fe/Cr interface with monoatomic 
steps from computer simulations assuming interdiffusion
over two layers. Note that the Fe moments are also affected.}
\label{spins}
\end{figure}

This phenomenological description is confirmed by computer simulations of a
classical Heisenberg model with the following Hamiltonian
\begin{equation}
  {\mathcal H} = -{1 \over 2}
  \sum_{\langle ij \rangle} {J_{e_i e_j} \vec s_i \cdot \vec s_j}
  - \sum_{i} {D_{e_i} (s_i^z)^2},
\end{equation}
where
\mbox{$\vec s_i = (s_i^x, s_i^y, s_i^z)$}
are spin vectors of unit length at site $i$ on a bcc(001) lattice, and
$e_i = \mathrm{Fe}, \mathrm{Cr}$
is the element at this site.
$J_{e_i e_j}$ is the nearest-neighbor exchange coupling constant
between elements $e_i$ and $e_j$, and $D_{e_i}$ is the uniaxial
anisotropy of element $e_i$ which, if positive, favors the $z$-direction
perpendicular to the film.
We assume that this anisotropy does not
depend on the position of the atoms.
We choose
$J_{\mathrm{Fe} \mathrm{Fe}}/k_B =  375K$ and
$J_{\mathrm{Cr} \mathrm{Cr}}/k_B = -170K$,
consistent with the critical
temperatures
$T^{\mathrm{com}} \approx 450K$ for Cr and
$T_{\mathrm{c}}   \approx 1000K$ for Fe in mean field solution of this model.
The interaction between Fe and Cr is chosen as
$J_{\mathrm{Fe} \mathrm{Cr}}/k_B = -40K$ \cite{highfield}.
The shape anisotropy of Fe induced by the dipole interaction is
modeled by a negative uniaxial anisotropy of
$D_{\mathrm{Fe}}/k_B = - 1.5K$.
Finally, we introduce a very small uniaxial anisotropy
$D_{\mathrm{Cr}}/k_B = 50mK$ at the Cr sites induced by epitaxial strain,
consistent with our results on uncovered Cr films \cite{Sonntag}.
Using a combination of over-relaxation dynamic and conjugate
gradient method, we determine the ground state of this system for
various configurations of the Fe/Cr interface.
This method is much faster than the tight binding method
used by Freyss et al.~\cite {StoefflerMRS}, while the magnetic structure
obtained is qualitatively the same.
To make our model even more realistic we have included interdiffusion at the
Fe/Cr interface \cite {FeCrexp} in addition to well defined steps \cite {yy}.
In Fig.~\ref{spins} the resulting ground state spin configuration
is shown.
Clearly, the
frustration induces an effective 90$^\circ$ coupling
between the Fe and the Cr order parameter (case 4). 
Together with the small
uniaxial anisotropy of the Cr atoms this leads to an orientation of
the Cr spins perpendicular to the surface, consistent with the experiment.

In our model we find this 90$^\circ$ orientation independent of the presence
of interdiffusion as long as there are steps.
Two length scales are important, the thickness of the Cr
layer and the separation of the steps at the interfaces. 
When $t_{\mathrm{Cr}}$ is reduced below the distance
between steps, more energy can be gained by the exchange interaction at the
interface than is lost by roughness induced domain wall formation within the
Cr.
Consequently, for thin Cr films the Cr moments are predicted to be oriented 
{\it in}
the film plane with domain walls in the Cr layer connecting the interfacial
steps.
Thus, our model also explains the observed reorientation transition with 
$t_{\mathrm{Cr}}$ (Fig.~\ref{phase2}) \cite {xx}.
For the thinnest Cr films the simulation yields a frustrated
spiral structure in the Cr, which induces strong non-collinear coupling
between the Fe layers in superlattices as predicted theoretically by
Slonczewski \cite {slonc} and confirmed experimentally by Schreyer et al.
\cite{Schreyer}.
Interestingly AF$_0$ Cr induces such coupling between the Fe layers
whereas SDW Cr does not \cite {Schreyer,Fullertonnew}.

Finally, we discuss the origin of the AF$_0$ phase.
For the smallest $t_{\mathrm{Cr}}$ well below the SDW period
no SDW can form (see Fig.~\ref{phase2}). Instead, the system becomes AF$_0$
\cite{Fullerton,Schreyer} consistent with theory \cite {fishman}.
Thus, the AF$_0$ order is induced by a finite size effect.
However, for thick films and in-plane propagation of the SDW, AF$_0$ order 
also occurs (see Figs.~\ref{data} and \ref{phase2}).
Grazing incidence X-ray and neutron experiments with depth resolved
information
have revealed that the in-plane SDW phase is located close to the
top Fe/Cr
interface \cite{oxford}. The AF$_0$ phase seems to be limited
to the lower Nb/Cr interface of the Fe/Cr/Nb/Sapphire structure.
The lower quality RHEED data of the Cr near the Nb interface mentioned
above indicates small {\it in-plane} crystalline grain dimensions near the
interface due to strain relaxation effects.
This can induce AF$_0$ order due to a finite size effect for the
in-plane SDW near the Cr/Nb interface.
With increasing $t_{\mathrm{Cr}}$ the {\it in-plane} crystalline quality improves
according to RHEED, allowing the formation of an in-plane SDW far away from the Cr/Nb
interface.
On the other hand, for out-of-plane SDW propagation
(Fig.~\ref{phase2})
no AF$_0$ phase occurs near the Nb/Cr interface, since in this case
the limiting factor is the {\it out-of-plane} crystalline grain size.
Using X-ray scattering we measure a much larger out-of-plane
coherence length than in the plane.
 
Consequently, a pure SDW propagating out-of-plane can form without any AF$_0$
contribution in the $t_{\mathrm{Cr}}$ range between 45 \AA \ and 250 \AA \ .
Thus, we can consistently explain the occurrence of the AF$_0$ phase by
finite size effects.

In conclusion, we have studied proximity effects between Fe and Cr
in the Fe/Cr system with neutron scattering.
We have focused on the regime of large $t_{\mathrm{Cr}}$ about
which no studies exist so far. As opposed to bulk Cr we find a single Q-state
SDW whose direction of propagation is reoriented from in-plane
to out-of-plane upon reducing $t_{\mathrm{Cr}}$. Compared to
bulk, the spin flip temperature is reduced to about 40 K.
The occurrence of commensurate AF$_0$ structures can be attributed to finite
size effects.
Using ground state calculations of a classical Heisenberg Hamiltonian the
observed reorientation transition is explained by a
realistic Fe/Cr interface with steps causing frustration of the system.

We thank Prof.~Usadel for
valuable discussions and J.~Podschwadek and
W.~Oswald for technical assistance.
The work in Bochum and Duisburg was supported
by the Deutsche Forschungsgemeinschaft through
SFB 166.


\begin{references}
\bibitem{highfield} This value was determined from a preliminary
high field magnetization measurement on an Fe/Cr superlattice.
\bibitem{yy} We use the same $J_{e_i e_j}$ and $D_{e_i}$
for the interdiffused region since deviations make no qualitative difference.
\end{references}
\end{document}